\documentclass[twocolumn]{aastex631}

\newcommand{\abund}[1]{$\log N({\rm #1})/N({\rm H})$}

\newcommand{\figref}[1]{Fig.~\ref{#1}}

\newcommand{\heone}{\ion{He}{1}}
\newcommand{\hetwo}{\ion{He}{2}}

\newcommand{\ctwo}{\ion{C}{2}}
\newcommand{\cthree}{\ion{C}{3}}
\newcommand{\cfour}{\ion{C}{4}}

\newcommand{\nthree}{\ion{N}{3}}
\newcommand{\nfour}{\ion{N}{4}}
\newcommand{\nfive}{\ion{N}{5}}

\newcommand{\othree}{\ion{O}{3}}
\newcommand{\ofour}{\ion{O}{4}}
\newcommand{\ofive}{\ion{O}{5}}
\newcommand{\osix}{\ion{O}{6}}

\newcommand{\sitwo}{\ion{Si}{2}}

\newcommand{\sifour}{\ion{Si}{4}}

\newcommand{\pfour}{\ion{P}{4}}
\newcommand{\pfive}{\ion{P}{5}}

\newcommand{\sfour}{\ion{S}{4}}
\newcommand{\sfive}{\ion{S}{5}}

\newcommand{\fefour}{\ion{Fe}{4}}
\newcommand{\fefive}{\ion{Fe}{5}}

\newcommand{\nifour}{\ion{Ni}{4}}
\newcommand{\nifive}{\ion{Ni}{5}}

\newcommand{\ebv}{$E(B-V)$\/}

\newcommand{\kms}{km s$^{-1}$}
\newcommand{\logg}{$\log g$}

\newcommand{\msun}{$M_{\sun}$}

\newcommand{\teff}{$T_{\rm eff}$}

\newcommand{\vlsr}{$v_{\rm LSR}$}

\shorttitle{III-60 in NGC 6723}
\shortauthors{Dixon}
\begin{document}

\title{Observations of the Ultraviolet-bright Star III-60 in the Globular Cluster NGC~6723}

\correspondingauthor{William V. Dixon}
\email{dixon@stsci.edu}

\author[0000-0001-9184-4716]{William V. Dixon}
\affiliation{Space Telescope Science Institute, 3700 San Martin Drive, Baltimore, MD 21218, USA}




\begin{abstract}

We have analyzed archival far-ultraviolet spectra of the UV-bright star III-60 in the globular cluster NGC~6723 obtained
with the Far Ultraviolet Spectroscopic Explorer (FUSE) and the  Cosmic Origins Spectrograph (COS).  We find that the star's photospheric parameters (effective temperature $T_{\rm eff} = 44{,}800  \pm 1200$, surface gravity $\log g = 4.89 \pm 0.18$, and helium abundance $\log N({\rm He})/N({\rm H}) = -0.84 \pm 0.29$) are consistent with the values derived from its optical spectrum, suggesting that optically-derived values are generally accurate for evolved stars with $T_{\rm eff} \lesssim$ 50,000 K.  Relative to the cluster's RGB stars, III-60 is enhanced in nitrogen and depleted in carbon and oxygen.  The star exhibits strong P~Cygni profiles in both components of the \ion{N}{5} $\lambda 1240$ doublet, but the resonance lines of other species show no evidence of a stellar wind.  The star's effective temperature and luminosity place it on the evolutionary tracks of stars evolving from the blue horizontal branch, but its high mass ($\sim 1.2 \, M_{\sun}$) indicates that it is the product of a stellar merger.  Its helium, carbon, and nitrogen abundances suggest that it is following an evolutionary path similar to that of the low-carbon, intermediate helium-rich hot subdwarfs.

%% The AAS Journals have a 250 word limit for the abstract.

\end{abstract}




\section{Introduction} \label{sec:intro}

 Among the UV-bright stars in the globular cluster NGC~6723---stars that lie above the horizontal branch and to the left of the giant branch in the Hertzsprung--Russell diagram---\citet{Menzies:1974} found two ``very blue stars,'' III-60\footnote{SIMBAD refers to this star as NGC 6723 360.  Its coordinates are 18:59:29.0 $-36$:40:49.0 \citep{Moehler:1998}.} and IV-9.  \citet{Moehler:1998} analyzed medium-resolution optical spectra of both stars and identified them as post-early AGB stars, objects that evolved partway up the asymptotic giant branch (AGB) before peeling off to the blue.  
 
\citet{Moehler:2019} reanalyzed the spectra, comparing them with grids of non-local thermodynamic equilibrium (non-LTE) line-blanketed model atmospheres and non-LTE synthetic spectra.  Using models with [M/H] = $-1.1$ and scaled-solar abundances, they derived \teff\ = $43{,}000 \pm 1400$ K, $\log g = 4.72 \pm 0.14$, and \abund{He} = $-1.19 \pm 0.14$ for the star III-60.  For many of the stars in their sample, \citeauthor{Moehler:2019}\ could not fit the H, \heone, and \hetwo\ lines simultaneously, because the observed \heone\ lines were weaker than predicted, while the H and \hetwo\ lines were stronger.  

\citet{Latour:2015} observed a similar behavior in the spectrum of the hot subdwarf O star BD+28\arcdeg 4211.  Suspecting that the stellar atmosphere was affected by opacities not included in the models, they found that model atmospheres with [M/H] = +1.0 were better able to fit the star's H and He lines and to reproduce the stellar parameters derived from far-ultraviolet (FUV) observations.  Using similar metal-enriched models, \citeauthor{Moehler:2019}\ derived \teff = $42{,}300 \pm 1400$, $\log g = 4.80 \pm 0.12$, and \abund{He} = $-1.13 \pm 0.12$ for III-60 in NGC~6723.  

Because the parameters derived from the [M/H] = $-1.1$ and +1.0 models agree within the uncertainties, we should have a good handle on the temperature of this star.  There is, however, one more consideration.  The spectrum of Y543 in NGC~6121 was also reanalyzed by \citet{Moehler:2019}.  Models with [M/H] = $-1.16$ yield \teff = $54{,}900 \pm 2000$, $\log g = 5.62 \pm 0.14$, and \abund{He} = $-1.25 \pm 0.10$, while those with [M/H] = +1.0 yield \teff = $56{,}500 \pm 1800$, $\log g = 5.71 \pm 0.12$, and \abund{He} = $-1.16 \pm 0.10$.  Again, the parameters agree within the uncertainties, but \citet{Dixon:2017} found that model fits to the CNO lines in the star's HST/COS spectrum require \teff\ $\sim$ 72,000 K, an effective temperature considerably higher than that derived from its optical spectrum.

To investigate whether a similar phenomenon is at work in III-60, we analyze archival FUV spectra spanning wavelengths from about 1765 \AA\ to the Lyman limit.  In Section \ref{sec_observations}, we present our data.  In Section \ref{sec_analysis}, we discuss our atmospheric models and use them to derive stellar parameters and abundances.  In Section \ref{sec_wind}, we discuss the wind profiles seen in the star's \ion{N}{5} $\lambda1240$ resonance lines.  In Section \ref{sec_discussion}, we discuss our results and consider the evolutionary history of the star.  We summarize our conclusions in Section \ref{sec_conclusions}.

%%%%%
% Table 1
\begin{deluxetable*}{lccccclcc}
\tablecaption{Summary of FUSE and COS Observations \label{tab:log_obs}}
\tablehead{
\colhead{Instrument} & \colhead{Grating} & \colhead{Cenwave} & \colhead{Wavelength} & \colhead{$R\equiv\lambda/\Delta\lambda$} & \colhead{Exp. Time} & \colhead{Obs.\ Date} & \colhead{Data ID} & \colhead{P.I.} \\
& & \colhead{(\AA)} & \colhead{Range (\AA)} & & \colhead{(s)}
}
\startdata
FUSE & $\cdots$ & $\cdots$ & \phn905--1187 & 20,000 &   \phn5,365 & 2000 May 15 & A1080201 & Dixon \\
           &                   &                      &                    &             &   13,537 & 2004 Sep 15 & D1570201 & Dixon \\
COS & G130M & 1291 & 1137--1433 & 15,000--21,000 & \phn\phn390 & 2010 Mar 4 & LB3Z11010 & Lehner \\
         & G130M & 1327 & 1174--1471 & 15,000--21,000 & \phn\phn310 & 2010 Mar 4 & LB3Z11020 & Lehner \\
         & G160M & 1577 & 1388--1754 & 15,000--22,000 & \phn\phn410 & 2010 Mar 4 & LB3Z11030 & Lehner \\
         & G160M & 1589 & 1400--1765 & 15,000--22,000 & \phn\phn490 & 2010 Mar 4 & LB3Z11040 & Lehner \\
\enddata
\end{deluxetable*}
%%%%%

\section{Observations and Data Reduction}\label{sec_observations}

The star III-60 was observed with both the Far Ultraviolet Spectroscopic Explorer (FUSE) and the Cosmic Origins Spectrograph (COS) aboard the {\em Hubble Space Telescope}.  Observational details are presented in Table \ref{tab:log_obs}. 

\subsection{FUSE Data}

FUSE provides medium-resolution spectroscopy from 1187 \AA\ to the Lyman limit \citep{Moos:00, Sahnow:00}.  The star III-60 was observed through the $30\arcsec \times 30\arcsec$ LWRS aperture.  The data were reduced using v3.2.2 of CalFUSE, the standard data-reduction pipeline software \citep{Dixon:07}, and retrieved from the Mikulski Archive for Space Telescopes (MAST).  For each FUSE channel, the extracted spectra from all exposures are shifted to a common wavelength scale, weighted by exposure time, and combined into a single file.  Because each channel has a unique line-spread function, we prefer not to combine the spectra from multiple channels, instead using only the spectrum from the channel with the highest signal-to-noise ratio (S/N).  In this case, in order to maximize the S/N, we combine spectra from the SiC1 and SiC2 channels for the wavelength region 917--988 \AA\ and from the LiF1 and LiF2 channels in the region 1180--1190 \AA.  We exclude the data from the SiC2 channel, exposures 1 and 2, of observation A1080201 and all of the Side 2 data of exposure D1570201, because the star had drifted out of the aperture.  The S/N of the resulting spectrum is $\sim 7$ per 0.05 \AA\ resolution element in the SiC band (900--1000 \AA) and ranges from 10 to 15 in the LiF band (1000--1187 \AA).  The FUSE wavelength calibration is reasonably accurate, but small offsets of the target within the LWRS aperture can introduce a zero-point offset in the wavelength scale.   To place the data on an absolute wavelength scale, we shift the spectrum so that the velocities of its interstellar lines match those of the COS spectrum, which are heliocentric.  The FUSE data used in this work are available at \dataset[10.17909/4jy0-pr10]{https://doi.org/10.17909/4jy0-pr10}.

\subsection{COS Spectroscopy}

COS enables high-sensitivity, medium- and low-resolution spectroscopy in the 1150--3200 \AA\ wavelength range \citep{Green:COS:2012}.   The star was observed with COS using the G130M and G160M gratings, each at multiple central-wavelength settings.  Because the CALCOS pipeline does not automatically combine exposures taken with different central wavelengths, we retrieved the fully-reduced and combined G130M and G160M spectra from the Hubble Advanced Spectral Product \citep[HASP;][]{Debes:2024} program, available from MAST.  The S/N per resolution element varies between 20 and 35 in the G130M spectrum.  In the G160M spectrum, the S/N falls monotonically from $\sim 35$ at 1400 \AA\ to $\sim 10$ at 1750 \AA.

\section{Analysis}\label{sec_analysis}

\subsection{Interstellar Medium}\label{sec_ism}

The FUV spectrum of III-60 includes a variety of interstellar absorption features.  In the FUSE bandpass, molecular hydrogen is the dominant species, while high-ionization species are prominent at longer wavelengths.  Synthetic interstellar absorption spectra are computed using software written at the University of California, Berkeley, by M. Hurwitz and V. Saba. Given the column density, Doppler broadening parameter, and velocity of each component, the program computes a Voigt profile for each absorption feature and produces a high-resolution spectrum of optical depth versus wavelength.  Wavelengths, oscillator strengths, and other atomic data are taken from \citet{Morton:03}.  We fit these features by eye, which is sufficient for our needs.  

\citet{Lehner:2012} report the detection of absorption from \ctwo\ $\lambda 1334$ and \sitwo\ $\lambda \lambda 1260, 1526$ with \vlsr\ $\sim \; -90$ \kms\ in the COS spectrum of III-60, which they attribute to a high-velocity cloud (HVC) along the line of sight.  Their estimate of the stellar velocity is $-101$ \kms, so stellar features are blue-shifted with respect to both the HVC and the local interstellar medium (ISM).  We have incorporated these features into our ISM model.  The question of whether any HVC absorption lines are blended with stellar features, complicating our estimate of the star's metal abundances, is addressed in Section \ref{sec_abundance}.

\subsection{Model Atmospheres}\label{sec_models}

We compute non-LTE stellar-atmosphere models using version 208 of the program TLUSTY \citep{Hubeny:Lanz:95}.  We employ atomic models similar to those used by \citet{Lanz:Hubeny:2003} to compute their grid of O-type stars.  We begin by adopting the stellar parameters derived by \citet{Moehler:2019}, \teff = $43{,}000 \pm 1400$, $\log g = 4.72 \pm 0.14$, and \abund{He} = $-1.19 \pm 0.14$, using models with [M/H] = $-1.1$ and scaled-solar abundances.  Given a model atmosphere, we compute a synthetic spectrum using version 54 of the program SYNSPEC \citep{Hubeny:88}.  For the FUSE data, synthetic spectra are convolved with a Gaussian of FWHM = 0.06 \AA\ to match the FUSE line-spread function.  For the COS spectrum, we employ the tabulated line-spread functions appropriate for data obtained at Lifetime Adjustment Position \#1, which are available from \href{http://www.stsci.edu/hst/cos/performance/spectral_resolution/}{the COS website}.  We multiply the synthetic spectrum by a \citet{Fitzpatrick:1999} extinction curve assuming \ebv\ = 0.05 \citep{Harris:96, Harris:2010}, extrapolated to the Lyman limit, and the ISM absorption model described above.  Finally, we scale the model to reproduce the continuum in a nearby (apparently) line-free region in the observed spectrum.

Given a grid of synthetic spectra, prepared as described above, our fitting routine linearly interpolates among them---in one, two, or three dimensions, as appropriate---determining the best fit to the data via chi-squared minimization.  The uncertainties quoted for parameters derived from individual line fits are $1 \sigma$ errors computed from the covariance matrix returned by the fitting routine; we refer to these as statistical errors.  

Continuum placement is the dominant uncertainty in our fits.  Weak absorption lines not included in our model may depress the apparent continuum.  Allowing our fitting routines to scale the model to the mean level of the ``pseudo-continuum'' thus underestimates the true continuum level.  To estimate the uncertainty inherent in our continuum estimate, we perform each fit twice, once with the model continuum fixed as described and again with the model scaled by a factor of 0.97.  The difference in the two abundances is an estimate of the systematic error in our abundance estimates. We add this term and the statistical error in quadrature to compute our final error for a single absorption feature.  In most cases, the continuum uncertainty is the dominant contributor to the final error.

\begin{figure}
\epsscale{1.15}
\plotone{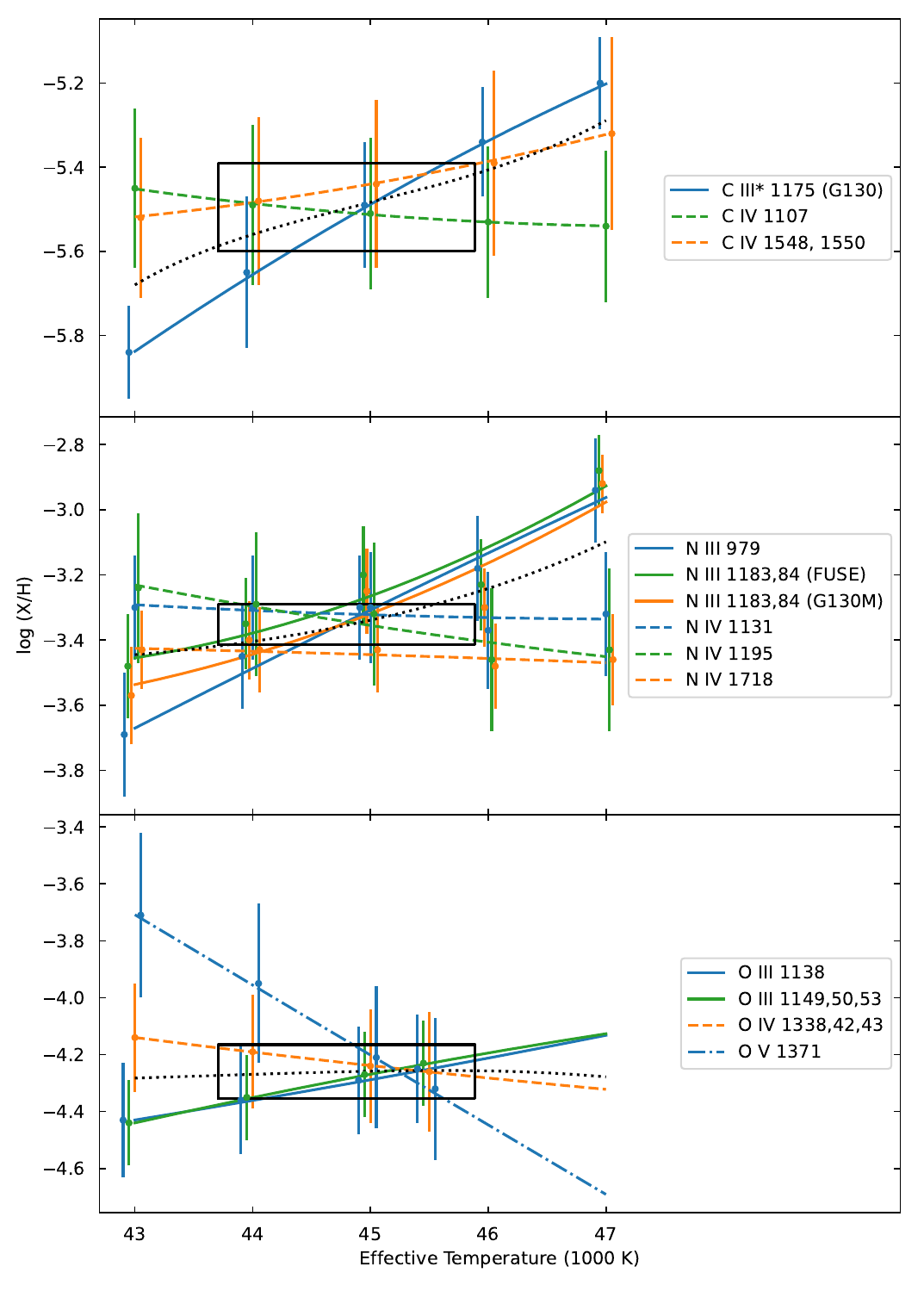}
\caption{Deriving the effective temperature and CNO abundance.  Points with error bars represent the abundance derived from model fits to each absorption feature.  Solid and dashed lines are low-order polynomial fits to the measured points.  Dotted lines represent the weighted mean of the measured abundances, computed at 10 K intervals.  The black box in each panel denotes the allowed ranges of temperature and abundance, as described in the text.}
\label{fig_teff}
\end{figure}

\subsection{Effective Temperature and CNO Abundances}\label{sec_teff}

Following \citet{Dixon:M5:2024}, we use the absorption features of multiple ionization states of CNO to derive the star's effective temperature.  Consider the middle panel of \figref{fig_teff}.  Using a series of models with \teff\ = 43,000 K and \abund{N} between $-4.5$ and $-2.5$, we fit the \nthree\ $\lambda 979$ feature to determine the nitrogen abundance.  We repeat using models with temperatures increasing in steps of 1000 K to 47,000 K.  The resulting nitrogen abundances are plotted as dark blue points and connected by a low-order polynomial (evaluated at 10 K intervals).  Vertical bars represent the uncertainties returned by the fitting routine.  As the temperature rises, the fraction of nitrogen in the form of \nthree\ falls, requiring a higher nitrogen abundance to reproduce the observed feature.  We repeat this procedure for a handful of other \nthree\ and \nfour\ lines.

We repeat the process for C and O (top and bottom panels of \figref{fig_teff}).  Note that the hottest available oxygen models have temperatures of 45,500 K; hotter models fail to converge.  We exclude from consideration features that yield wildly discrepant abundances, as well as those that yield abundances with error bars so large that they provide no useful constraints.  We see that the curves cross at a temperature near 45,000 K.  To combine these results in a quantifiable way, we compute the error-weighted mean abundance and the error-weighted standard deviation as a function of temperature for each element.  The mean abundance is plotted as a dotted line in each panel.  At each temperature step, we compute
$$\chi^2 = \sum \Bigl \lbrace [y_i - y(x_i)]^2 / \sigma_i^2 \Bigr \rbrace,$$
where $y_i$ is the abundance derived from a single feature (or group of features; solid or dashed line), $y(x_i)$ is the mean abundance (dotted line), $\sigma_i$ is the uncertainty in the derived abundance (vertical bars), and the summation is taken over the 13 abundance curves plotted in \figref{fig_teff}.  $\chi^2$ has a minimum at \teff\ = 44,840 K.  Our lower limit to the temperature is \teff\ = 43,710 K, set by the point at which $\chi^2$ rises by 4.72 relative to its minimum.  This choice of $\Delta \chi^2$ is strictly correct for a model with four interesting parameters (one temperature and three abundances) if all errors are normally distributed \citep{Press:89}.  For each element, our best-fit abundance is the error-weighted mean value (computed above) at the best-fit temperature.  The abundance uncertainty is the larger of the error-weighted standard deviation or the uncertainty in the weighted mean.  The black box in each panel illustrates the best-fit effective temperature and elemental abundance and their uncertainties.  Our derived stellar parameters are presented in Table \ref{tab:stellar_parms} and abundances in Table \ref{tab:abundance}.

\subsection{Surface Gravity and Helium Abundance}\label{sec_gravity}

\begin{figure}
\epsscale{1.15}
\plotone{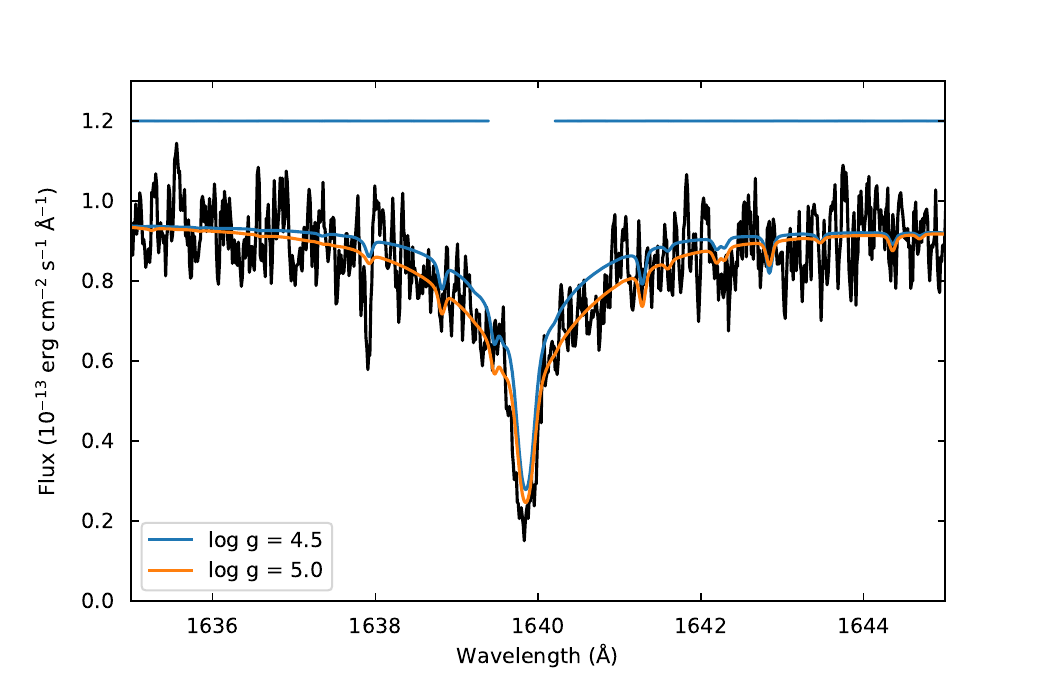}
\caption{\hetwo\ $\lambda 1640$ feature in the spectrum of III-60 in NGC~6723.  For this figure, the data are smoothed by three pixels.  Synthetic spectra with \logg\ = 4.5 and 5.0 are overplotted.  Both models have \teff\ = 44,800 K and \abund{He} = $-0.84$.  Blue bars indicate the spectral regions included in our fit.}
\label{fig_1640}
\end{figure}

To determine the surface gravity and helium abundance of III-60, we fit a small grid of models to the star's \hetwo\ $\lambda 1640$ line.  The models have \teff\ = 44,800 K; \logg\ = 4.50, 4.75, and 5.00; and \abund{He} = $-0.5$, $-1.0$, and $-1.5$.  Our best-fit model has \logg\ = $4.89 \pm 0.18$ and \abund{He} = $-0.84 \pm 0.29$.  The \hetwo\ feature, along with two of our models, is presented in \figref{fig_1640}.  We see that increasing the surface gravity increases the depth of the \hetwo\ line.  Increasing the He abundance has the same effect, so we cannot place tight constraints on the two parameters simultaneously, as indicated by their relatively large error bars.  Note that the line core, which is both broader and deeper than our models, is excluded from the fit.  

Formally, our best-fit surface gravity and helium abundance are consistent with the values $\log g = 4.80 \pm 0.12$ and \abund{He} = $-1.13 \pm 0.12$ determined from the star's optical spectrum \citep{Moehler:2019}; however, the fact that we derive larger values for both parameters suggests that the \hetwo\ $\lambda 1640$ line is stronger than its optical counterparts.  Indeed, fixing the surface gravity at its optically-derived value yields a helium abundance \abund{He} = $-0.69$.  The \hetwo\ $\lambda 958$ line is also quite strong, but its low S/N precludes its use in this analysis.  In the rest of this paper, we will adopt the FUV-derived values of the surface gravity and helium abundance while keeping in mind the uncertainties in both parameters.

Are the new helium abundance and surface gravity sufficiently different from our initial models that we must recompute the temperature and CNO abundance?  To find out, we generate a grid of models with \teff\ = 45,000 K, \logg\ = 4.89,  \abund{He} = $-0.84$, and \abund{N} ranging from $-2.5$ to $-4.5$.  Fitting this grid to our set of nitrogen lines yields \abund{N} = $-3.35 \pm 0.04$.  Fits to our initial set of models with the same temperature yields \abund{N} = $-3.30 \pm 0.07$.  The difference is small enough that we need not revisit our earlier computations.

%%%%%
% Table 2
\begin{deluxetable}{lcccc}[t]
\tablecaption{Stellar Parameters \label{tab:stellar_parms}}
\tablehead{
\colhead{Parameter} & \colhead{Value}
}
\startdata
\teff\ (K) & $44{,}800  \pm 1100$ \\
\logg\ [cm s$^{-2}$] & $4.89 \pm 0.18$ \\
\abund{He} & $-0.84 \pm 0.29$ \\
$R_*/R_{\sun}$ & $0.65 \pm 0.01$ \\
$M_*/M_{\sun}$ & $1.18 \pm 0.49$ \\
$\log (L_*/L_{\sun})$ & $3.18 \pm 0.05$ \\
\enddata
\end{deluxetable}
%%%%%

\subsection{Metal Abundances}\label{sec_abundance}

As mentioned above, some of the star's CNO lines yield abundances that are discrepant from others of their species or whose error bars are so large that they provide no constraint on the effective temperature.  To give a sense of these features, we present in Table~\ref{tab:lines_cno}, found in the Appendix, abundances derived from a number of CNO features, including several that are not included in \figref{fig_teff}.

\begin{figure}
\epsscale{1.15}
\plotone{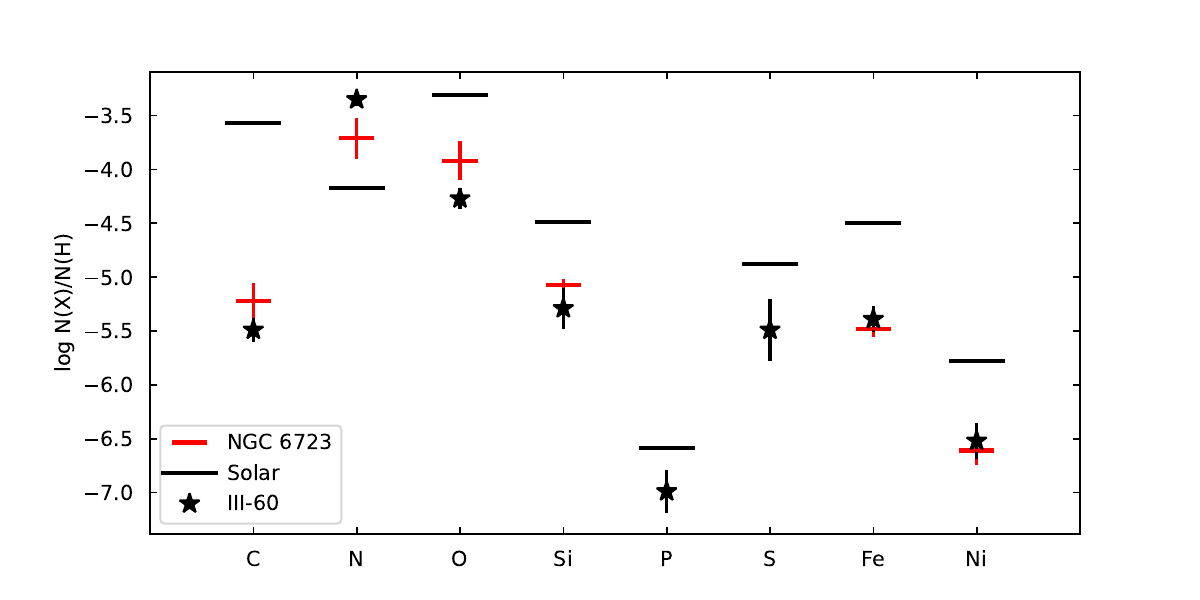}
\caption{Photospheric abundances of III-60 (stars), NGC~6723 (red lines), and the solar photosphere (black lines).  See Table \ref{tab:abundance} for references.}
\label{fig_abundance}
\end{figure}

Five other elements, Si, P, S, Fe, and Ni, exhibit absorption lines in the spectrum of III-60.  We derive their abundances by fitting synthetic spectra to the features listed in Table~\ref{tab:lines_fuv}, also found in the Appendix.  We fit each line or group of lines separately.   For each of these elements, the final abundance and its uncertainly represent the mean and standard deviation of our individual measurements.  Results are presented in Table \ref{tab:abundance} and plotted in \figref{fig_abundance}.  

Notes on individual elements follow.

{\em Carbon:}\/ The two components of the \cfour\ $\lambda \lambda 1548, 1550$  resonance doublet are strong in both the stellar photosphere and the local ISM.  If significant \cfour\ were present in the HVC reported by \citet{Lehner:2012}, its features would be blended with the corresponding stellar lines.  To minimize interstellar contamination, we fit only the blue wing of each component of the doublet.  These lines yield carbon abundances similar to those of the star's other \cfour\ features, suggesting that any contamination is minor.

{\em Nitrogen:}\/  The \nfour\ $\lambda 1718$ feature is quite strong, but its blue wing is blended with a couple of stellar features, probably due to \fefive, that are not reproduced by our model.  We fit only the red wing of the line.  We do not attempt to fit the \nfive\ $\lambda \lambda 1238, 1242$ resonance doublet, as both components are hopelessly entangled with the P~Cygni profiles discussed in Section \ref{sec_wind}.  

{\em Silicon:}\/ The discussion of the \cfour\ $\lambda \lambda 1548, 1550$ lines also applies to the \sifour\ $\lambda \lambda 1393, 1402$ resonance doublet.

{\em Phosphorus:}\/ If significant \pfive\ were present in the HVC, then the stellar resonance features at 1117 and 1128 \AA\ could be contaminated.  Furthermore, our models predict that \pfour\ $\lambda 1033$ is blended with an \othree\ feature, while \pfive\ $\lambda 1117$ is blended with \nfour.  Nevertheless, the scatter in the abundance values derived from the four P lines listed in Table \ref{tab:abundance} is relatively small, so we adopt their mean as the stellar value.

{\em Iron:}\/  There are hundreds of iron lines in the FUV spectrum of III-60, most due to \fefive.  Instead of fitting individual features, we divide the G160M spectrum into several short segments, mask the ISM features and all stellar features that are not due to iron, and fit our model spectra to each segment.

{\em Nickel:}\/ In most cases, we fit individual nickel lines.  One exception is the region 1265--1268 \AA, which contains several strong \nifive\ features.  We fit our model spectra to this segment, just as we did for the iron lines.

%%%%%
% Table 3
\begin{deluxetable}{lccc}
\caption{Photospheric Abundances \label{tab:abundance}}
\tablehead{
\colhead{Species} & \colhead{III-60} & \colhead{NGC 6723} & \colhead{Sun}
}
\startdata
He & $-0.84 \pm 0.29$ & \nodata & $-1.07 \pm 0.01$ \\
C  & $-5.50 \pm 0.10$ & $-5.22 \pm 0.17$ & $-3.57 \pm 0.05$ \\
N  & $-3.35 \pm 0.06$ & $-3.71 \pm 0.19$ & $-4.17 \pm 0.05$ \\
O  & $-4.26 \pm 0.09$ & $-3.92 \pm 0.18$ & $-3.31 \pm 0.05$ \\
Si & $-5.29 \pm 0.19$ & $-5.07 \pm 0.05$ & $-4.49 \pm 0.03$ \\
P  & $-6.99 \pm 0.20$ & \nodata & $-6.59 \pm 0.03$ \\
S  & $-5.49 \pm 0.29$ & \nodata & $-4.88 \pm 0.03$ \\
Fe & $-5.39 \pm 0.12$ & $-5.48 \pm 0.08$ & $-4.50 \pm 0.04$ \\
Ni & $-6.52 \pm 0.17$ & $-6.61 \pm 0.13$ & $-5.78 \pm 0.04$ \\
\enddata
\tablecomments{Abundances relative to hydrogen: \abund{X}.  Cluster values for C and N from \citet{Gerber:2019}, for O, Si, and Fe from \citet{Rojas-Arriagada:2016}, and for Ni from \citet{Crestani:2019}.  Solar values from \citet{Asplund:2009}.}
\end{deluxetable}
%%%%%

\section{Stellar Wind}\label{sec_wind}

\begin{figure}
\epsscale{1.15}
\plotone{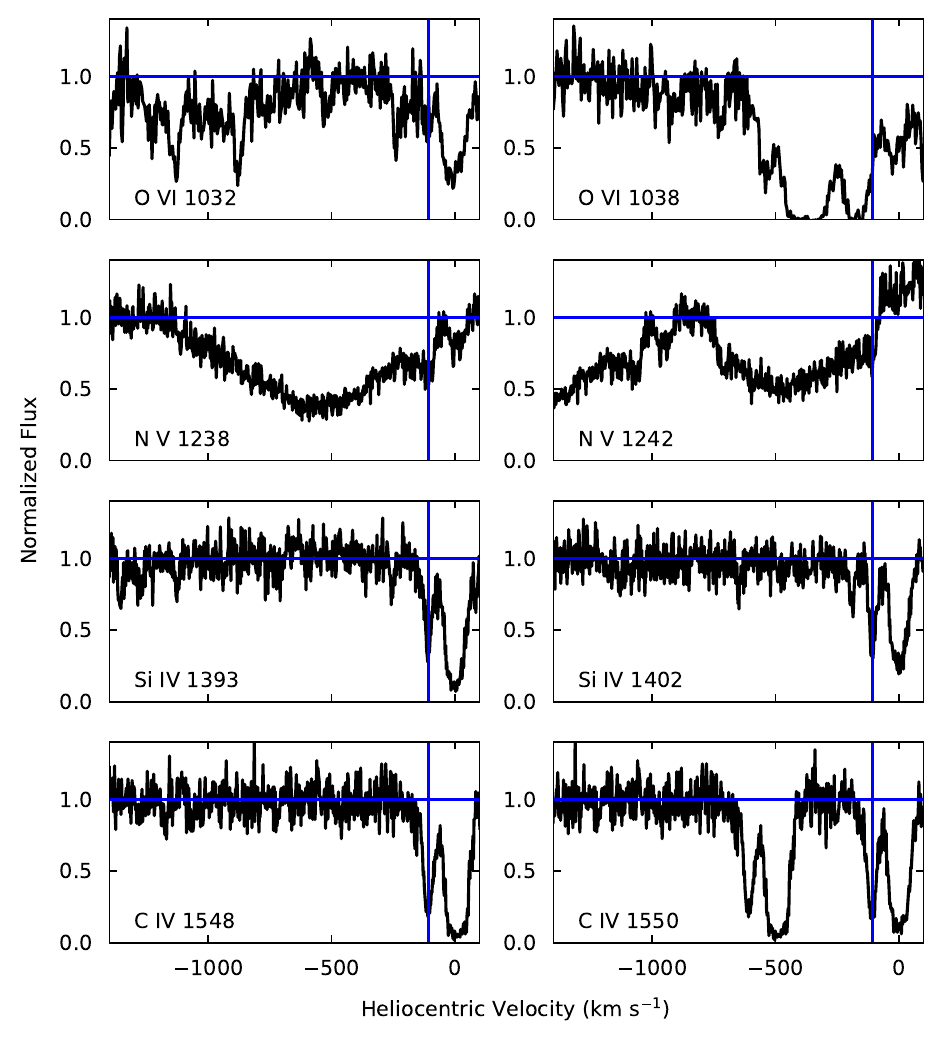}
\caption{Normalized profiles of the resonance doublets observed in the spectrum of III-60. The horizontal line corresponds to the normalized continuum. The vertical line indicates the radial velocity of the star.  Except for the resonance lines, all of the features are interstellar.}
\label{fig_doublets}
\end{figure}

The FUV spectra of many hot post-AGB stars exhibit wind features from a variety of species \citep[e.g.,][]{Guerrero:DeMarco:2013}.  \figref{fig_doublets} presents the resonance lines of \ion{O}{6} $\lambda1034$, \ion{N}{5} $\lambda1240$, \ion{Si}{4} $\lambda 1397$, and \ion{C}{4} $\lambda1550$.  Only the \ion{N}{5} $\lambda 1240$ doublet shows evidence of a wind. Its blue component exhibits a P~Cygni absorption trough extending about $1100$ km~s$^{-1}$ blueward of the photospheric line and reaching depths of about 60\% below the continuum. Its red component exhibits red-shifted emission.  The continuum for the other resonance lines is flat; no broad absorption is observed.

\citet{Chayer:2015} found that, like III-60 in NGC~6723, the UV-bright star vZ~1128 in M3 (NGC~5272) exhibits P~Cygni profiles in both components of the \nfive\ $\lambda 1240$ doublet, but in no other absorption lines.  They performed a simple calculation to show that the low abundance of C and Si (relative to N), combined with the tiny fraction of these elements in the form of \cfour\ and \sifour, suggest that the optical depth of their wind features would be only a few percent that of \nfive.  The same argument explains the absence of \osix, \sfour, and \cfour\ wind lines in the spectrum of III-60.

\section{Discussion}\label{sec_discussion}

\subsection{Cluster Membership}\label{sec_membership}

Before comparing III-60 with other members of NGC~6723, we should confirm that it is a cluster member.  The star lies about 3$\arcmin$ from the cluster center \citep{Menzies:1974}, and its radial velocity, \vlsr\ $= -101 \pm 10$ \kms\  \citep{Lehner:2012}, is consistent with the cluster mean of \vlsr\ $= -88.3 \pm 3.6$ \kms\ \citep{Harris:96, Harris:2010}.  \citet{Vasiliev:2021} use data from {\em Gaia} Early Data Release 3 \citep[EDR3;][]{Gaia_Mission, GaiaEDR3} to study the kinematic properties of 170 Milky Way globular clusters.  They provide catalogues of all {\em Gaia}\/ sources in the field of each cluster.  They estimate a membership probability of 0.997 for III-60.  We conclude that the star is a cluster member.

\subsection{Photospheric Abundances}\label{sec_abundance_discussion}

Table \ref{tab:abundance} and \figref{fig_abundance} compare the measured abundances of III-60 with the average of the cluster's RGB stars and with the sun.  The star's He abundance is super-solar, but the uncertainty is large.  The star is enhanced in N and depleted in C and O, a pattern generally attributed to hydrogen burning via the CNO cycle.  The star's Si, Fe, and Ni abundances are consistent with the cluster values.  Any evolutionary scenario proposed for III-60 must be able to explain this abundance pattern.  Note that we have adopted the Fe abundance of \citet{Rojas-Arriagada:2016}, who found [Fe/H] = $-0.98 \pm 0.08$, but the cluster abundance is somewhat uncertain.  Recent spectroscopic estimates range from [Fe/H] = $-0.93 \pm 0.05$ \citep{Crestani:2019} to [Fe/H] = $-1.22 \pm 0.08$ \citep{Gratton:2015}.

\subsection{Stellar Mass and Luminosity}\label{sec_mass}

We can derive a star's radius, and from this its mass and luminosity, by comparing its observed and predicted fluxes.  The spectral irradiance of III-60, as reported in {\em Gaia} Data Release 3 \citep[DR3;][]{Gaia_Mission, GaiaDR3}, is $G = 15.552 \pm 0.003$ mag.  (The star is designated Gaia DR3 6730904559676428544.)  We generate a synthetic spectrum using our best-fit model, scaled by a \citet{Fitzpatrick:1999} extinction curve with $R_V = 3.1$ and \ebv\ = 0.05 \citep{Harris:96, Harris:2010}.  We compute synthetic stellar magnitudes using the recipe provided in \citet{Riello:2020}.  The ratio between the observed and model fluxes is $\phi = (3.89 \pm 0.01) \times 10^{-23}$.

In the synthetic spectra generated by SYNSPEC, the flux is expressed in terms of the flux moment, $H_\lambda$.  If the star's radius and distance are known, then the scale factor required to convert the model spectrum to the flux at earth is $\phi = 4 \pi (R_* / d)^2$ \citep{Kurucz:79}.  Using a combination of Gaia EDR3, {\it HST}, and literature data, \citet{Baumgardt:2021} derive a distance to NGG~6723 of $8.267 \pm 0.100$ kpc.  Adopting this distance and our scale factor, we derive a stellar radius  $R_*/R_{\sun}$ of $0.65 \pm 0.01$.  Applying our adopted surface gravity (\logg\ = $4.89 \pm 0.18$), we find that the stellar mass $M_*/M_{\sun}$ is $1.18 \pm 0.49$.  Finally, combining the stellar radius with our best-fit effective temperature (\teff\ = $44{,}800 \pm 1100$ K), we derive a stellar luminosity $\log L_*/L_{\sun}$ of $3.18 \pm 0.05$.  (The uncertainties on these values are calculated via simple propagation of errors.)

While a stellar mass of 1.18 \msun\ greatly exceeds the $0.53 \pm 0.01 \, M_{\sun}$ that one would expect for a soon-to-be white dwarf in a globular cluster \citep{Kalirai:2009}, the value seems secure.  Adopting the \citet{Gontcharov:2023} correction for differential reddening across the cluster yields \ebv\ = 0.069 and $M_*/M_{\sun} = 1.25$.  Adopting the scale factor from our fits to the star's G160M spectrum ($\phi = 3.31 \times 10^{-23}$) yields $M_*/M_{\sun} = 1.00$.  From the star's optical spectrum, \citet{Moehler:2019} derived a mass of $1.19 \pm 0.3$ \msun\ (the error bar is from their Fig. 5).

\subsection{Opacity Effects}\label{sec_opacity}

One motivation of this study is to determine whether the star's optical hydrogen and helium features yield stellar parameters, particularly effective temperature, consistent with those derived from the ionization balance of the star's FUV lines.  In this case, the answer is yes, but careful consideration of our results provides interesting insights into opacity effects in stellar atmospheres.

The discrepancy between the optical and UV-derived effective temperatures may be a symptom of the Balmer-line problem \citep{Napiwotzki:93, Werner:96}, the inability of models to reproduce simultaneously the full set of Balmer lines with a single set of stellar parameters (\teff\ and \logg).  Apparently, there are sources of opacity in the atmospheres of hot stars that are not included in our models.  For only a handful of globular-cluster stars are both optical and UV temperatures available.  We do not find a significant discrepancy in III-60, for which \teff\ $\sim$ 45,000 K.  \citet{Latour:2017} found that, while the optically-derived temperatures of subdwarf O (sdO) stars in $\omega$ Cen are about 50,000 K, the UV-derived temperatures are closer to 60,000 K.  The UV-bright star Y453 in M4 (NGC~6121) yields an optical temperature of 56,500 K \citep{Moehler:2019} and a UV temperature of 72,000 K \citep{Dixon:2017}.  If effective temperature is the only relevant parameter (which is unlikely), then the discrepancy appears in stars hotter than about 50,000 K.

Another source of opacity is the diffusion of heavy elements into the stellar atmosphere.  The abundance ratios of these elements represent an equilibrium among several physical processes, including radiative levitation, which injects heavy elements into the atmosphere from below; gravitational settling, which pulls them from the atmosphere into the interior; and the stellar wind, which ejects them into the ISM.

The abundance anomalies seen in some sdO stars (which have \teff\  $>$ 40,000 K and \logg\ = 5.0--6.5), particularly enrichments in iron-group and trans-iron elements, are attributed to diffusion processes in the stellar atmosphere \citep{Heber:2016}.  We have seen that the photospheric abundances of III-60 are consistent with the cluster values (though its CNO abundances have clearly been modified).  With stellar parameters similar to those of sdO stars, III-60 might show abundance anomalies if not for its stellar wind.  Diffusion velocities in stellar atmospheres may reach about 1 cm s$^{-1}$ in the line-forming region.  When the wind velocity is significantly greater than the diffusion velocity, the accumulation of material lifted from below is impossible \citep[][pp. 134--135]{Michaud:2015}.  We can compare III-60 with the Y453, whose abundances are clearly influenced by diffusion processes; that star has \teff\ = 72,000 K, \logg\ = 5.7, and no detectible wind \citep{Dixon:2017}.

\subsection{Evolutionary Status}\label{sec_evolution}

The effective temperature and luminosity of III-60 place it on the evolutionary tracks of stars evolving from the blue horizontal branch \citep[BHB;][]{Moehler:2019}.  NGC~6723 has an extended HB, well populated on both sides of the RR Lyrae instability strip, so a BHB origin for III-60 would be reasonable if its mass were roughly half the measured value.  Instead, the star's high mass suggests that III-60 is the product of a stellar merger.  We consider several possibilities.

The simplest explanation is that the star is a blue straggler, still on the main sequence, but a main-sequence star of this temperature would have a mass of 120 \msun\ and a radius of 15 $R_{\sun}$ \citep{Cox:2000}.  We can also exclude the possibility that the star is a blue straggler that evolved through the AGB stage without undergoing further interactions with its neighbors.  Such stars might have main-sequence masses of 2 or 3 \msun, but the stellar-evolution models of  \citet{Miller_Bertolami:2016} predict that they evolve into objects with final masses of 0.6 to 0.8 \msun, lower than is observed.  Furthermore, in the post-AGB phase, these stars are enhanced in carbon and depleted in nitrogen (a result of third dredge-up on the AGB), while III-60 is depleted in carbon and enhanced in nitrogen.

\begin{figure}
\epsscale{1.15}
\plotone{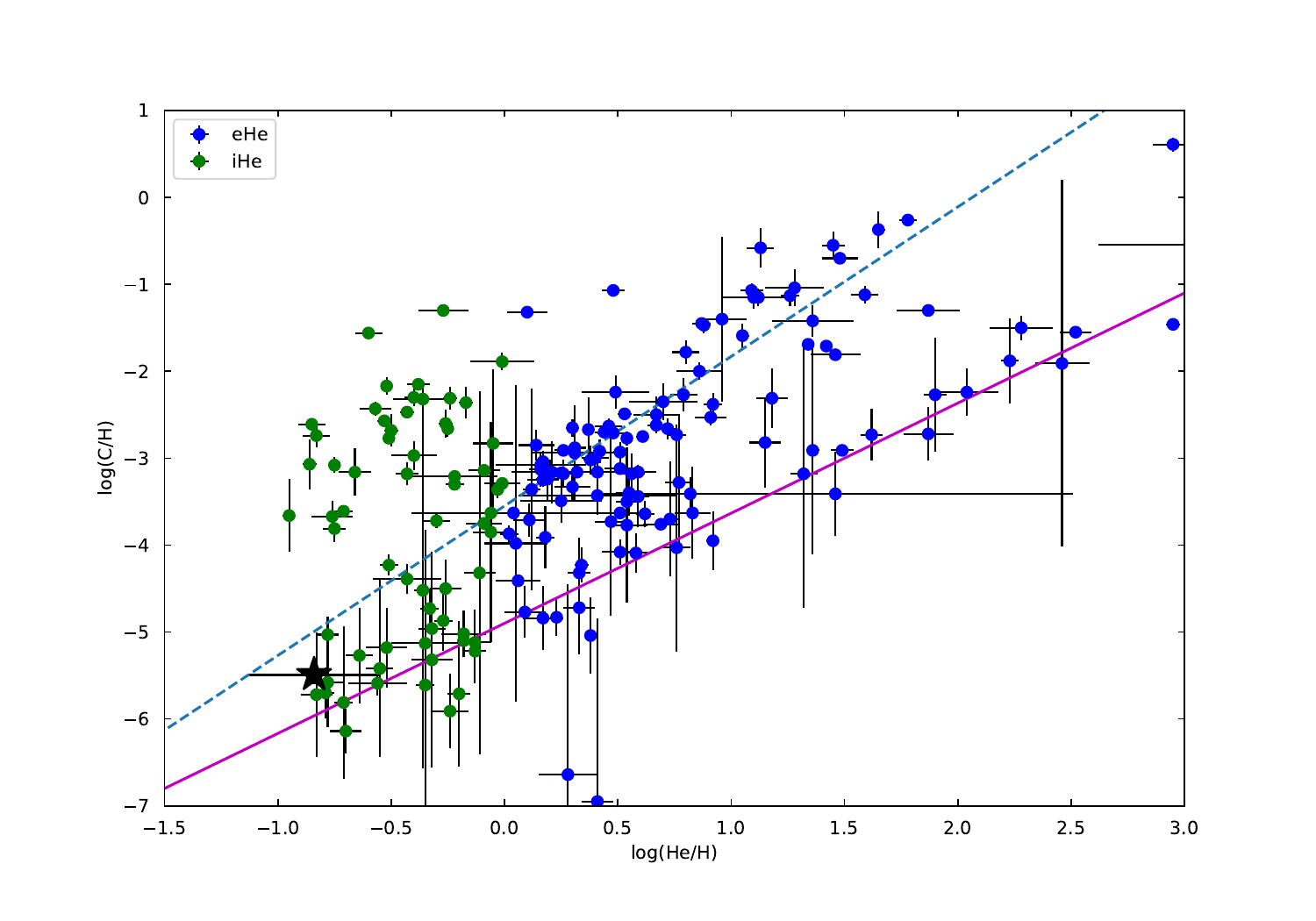}
\plotone{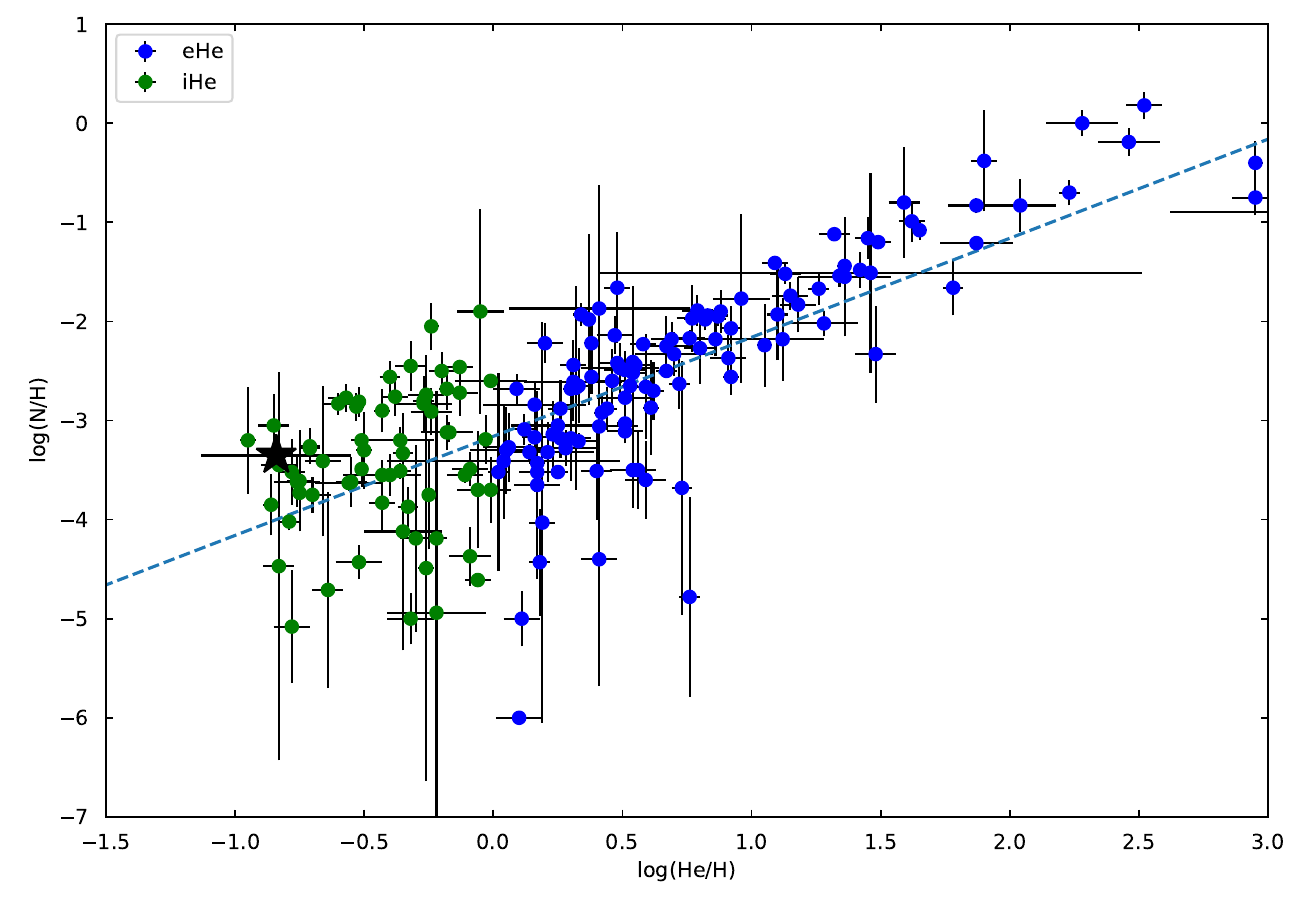}
\caption{Abundance correlations of hot subdwarf stars. Top panel:  carbon abundance vs.\ helium abundance. Bottom panel:  nitrogen abundance vs.\ helium abundance. Extreme helium-rich (eHe) stars are plotted in blue, intermediate helium-rich (iHe) stars in green.  III-60 is in black.  Stellar data are from \citet{Luo:2024}.  Dashed lines represent the best-fit trends of \citet{Nemeth:2012}.  Solid line is a trend identified by \citeauthor{Luo:2024}}
\label{fig_carbon}
\end{figure}

III-60 must be the result of a stellar merger that occurred at a later evolutionary stage.  Perhaps the star is related to the hot subdwarf stars, some of which may be the products of the merger of two helium white dwarfs \citep{Webbink:84} or of a helium white dwarf and a low-mass main-sequence companion \citep{Clausen:2011}.  \citet{Luo:2024} derived the abundances of He, C, and N in 210 He-rich hot subdwarfs included in both the Gaia DR3 and LAMOST DR7 \citep{Luo:2022} data sets.  They distinguished between extreme helium-rich (eHe) stars, with \abund{He} $> 0$, and intermediate helium-rich (iHe) stars, with $-1 < $ \abund{He} $< 0$.  \figref{fig_carbon}, reproduced from their paper, shows correlations between the carbon (upper panel) and nitrogen abundance (lower panel) and the helium abundance.  In both panels, the dashed lines represents the trends identified by \citet{Nemeth:2012}.  

\citet{Luo:2024} identified three groups of stars in the C-He plane: eHe stars (blue points) that follow the \citet{Nemeth:2012} relation, iHe stars (green points) that lie above the \citeauthor{Nemeth:2012}\ relation, and iHe and eHe stars that fall on a shallower relation, marked in \figref{fig_carbon} with a solid line.  \citeauthor{Luo:2024} suggested that these three groups may represent three different mechanisms for the production of hot subdwarfs.  In the N-He plane, all of the stars fall on the \citeauthor{Nemeth:2012}\ relation.  The abundances of III-60 (black star) place it among the low-carbon iHe stars.

\begin{figure}
\epsscale{1.15}
\plotone{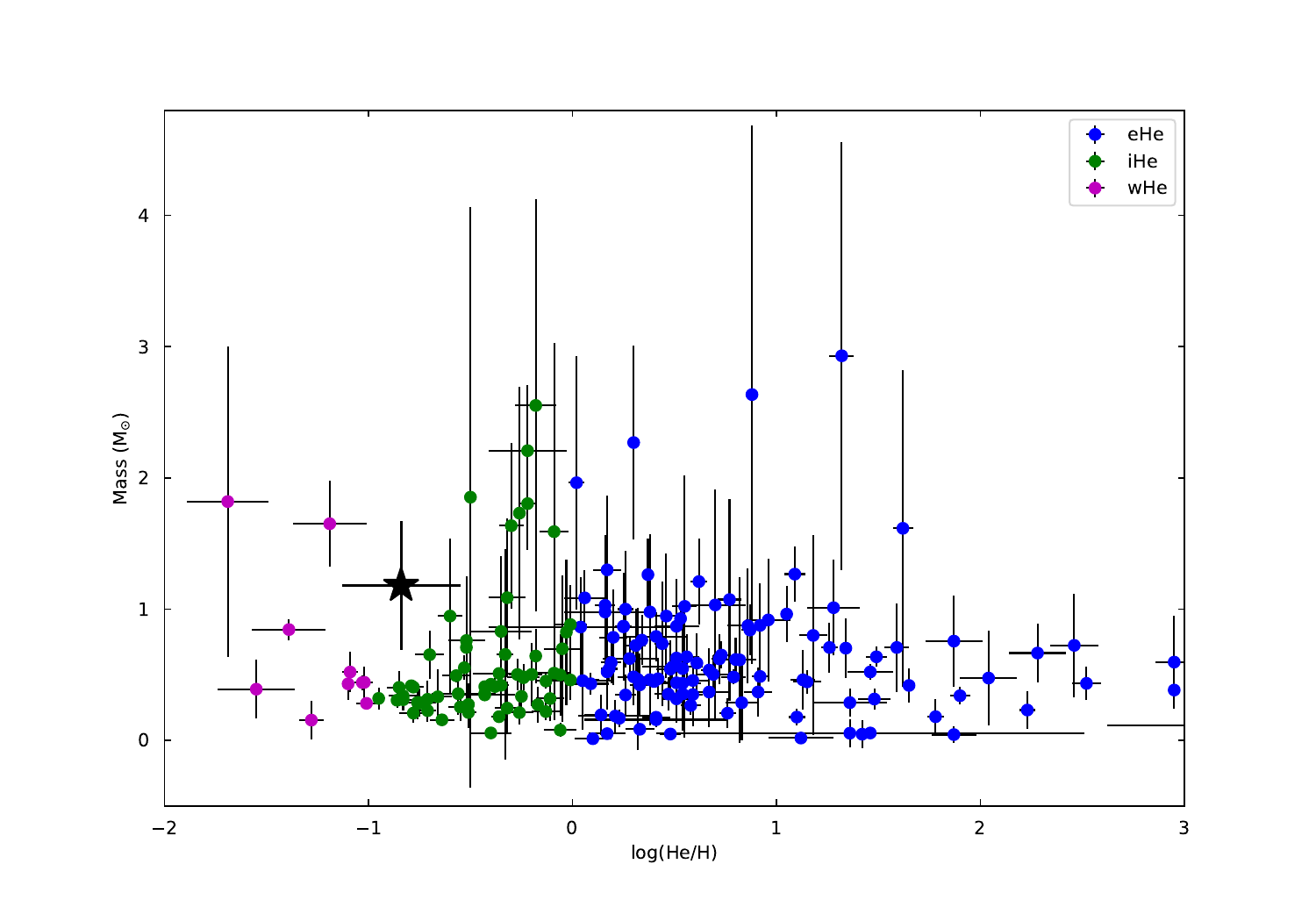}
\caption{Stellar mass vs.\ helium abundance. Extreme helium-rich (eHe) stars are plotted in blue, intermediate helium-rich (iHe) stars in green, and helium-weak (wHe) stars in magenta.  III-60 is black.  Stellar data are from \citet{Luo:2024}.}
\label{fig_mass}
\end{figure}

How does the mass of III-60 compare with those of the hot subdwarfs?  In \figref{fig_mass}, we plot stellar mass as a function of helium abundance.  In this figure, we include the helium-weak stars (wHe; magenta points), with $-2.2 < $ \abund{He} $< -1.0$, from the \citet{Luo:2024} sample.  We see that, while most hot subdwarfs have masses near 0.5 \msun, masses near 1 \msun\ are not unusual, even among stars with sub-solar He abundance.  We conclude that III-60 may be the product of an evolutionary path similar to that of the low-carbon iHe hot subdwarfs.

Note: We derive stellar masses for the \citet{Luo:2024} sample by converting their Gaia G-band magnitudes to V-band magnitudes using the relation provided in Table 5.9 of the \href{https://gea.esac.esa.int/archive/documentation/GDR3/Data_processing/chap_cu5pho/cu5pho_sec_photSystem/cu5pho_ssec_photRelations.html}{Gaia DR3 Documentation}.  We convert from observed to bolometric magnitudes using the dust maps of the \citet{Planck:2014} and the bolometric corrections of \citet{Flower:1996} as amended by \citet{Torres:2010}.  The distance, effective temperature, and surface gravity of each star are taken from the \citeauthor{Luo:2024} catalog.

\section{Conclusions}\label{sec_conclusions}
 
We have analyzed archival FUV spectra of the hot UV-bright star III-60 in the globular cluster NGC~6723.  We find that the star's photospheric parameters (effective temperature, surface gravity, and helium abundance) are consistent with the values derived from its optical spectrum, suggesting that optically-derived values are generally accurate for evolved stars with \teff\ $\lesssim$ 50,000 K.  Relative to the cluster's RGB stars, III-60 is enhanced in nitrogen and depleted in carbon and oxygen.  The star exhibits strong P~Cygni profiles in both components of the \ion{N}{5} $\lambda 1240$ doublet, but the resonance lines of other species show no evidence of a stellar wind.  The star's effective temperature and luminosity place it on the evolutionary tracks of stars evolving from the BHB, but its high mass suggests that it is the product of a stellar merger.  Its helium, carbon, and nitrogen abundances suggest that it is traversing an evolutionary path similar to that of the low-carbon, intermediate helium-rich hot subdwarfs.

\begin{acknowledgments}

The author would like to thank P. Chayer for many helpful discussions.
This work has made use of
NASA's Astrophysics Data System (ADS); % and
the SIMBAD database, operated at CDS, Strasbourg, France; % and
%
% the NASA/IPAC Extragalactic Database (NED), which is operated by the Jet Propulsion Laboratory, California Institute of Technology, under contract with the National Aeronautics and Space Administration; and
%
the Mikulski Archive for Space Telescopes (MAST), hosted at the Space Telescope Science Institute, which is operated by the Association of Universities for Research in Astronomy, Inc., under NASA contract NAS5-26555; and
%
%data, software and/or web tools obtained from the High Energy Astrophysics Science Archive Research Center (HEASARC), a service of the Astrophysics Science Division at NASA/GSFC and of the Smithsonian Astrophysical Observatory's High Energy Astrophysics Division; and
%
% the Keck Observatory Archive (KOA), which is operated by the W.\ M.\ Keck Observatory and the NASA Exoplanet Science Institute (NExScI), under contract with the National Aeronautics and Space Administration; and
%
% the Two Micron All Sky Survey, which is a joint project of the University of Massachusetts and the Infrared Processing and Analysis Center/California Institute of Technology, funded by the National Aeronautics and Space Administration and the National Science Foundation; and
%
% the Wide-field Infrared Survey Explorer, which is a joint project of the University of California, Los Angeles, and the Jet Propulsion Laboratory/California Institute of Technology, funded by the National Aeronautics and Space Administration; and
%
data from the European Space Agency (ESA) mission
Gaia (\url{https://www.cosmos.esa.int/gaia}), processed by the Gaia
Data Processing and Analysis Consortium (DPAC;
\url{https://www.cosmos.esa.int/web/gaia/dpac/consortium}). 
Funding for the DPAC has been provided by national institutions, 
in particular the institutions participating in the Gaia Multilateral Agreement.
Publication of this work is supported by the STScI Director's Discretionary Research Fund.

\end{acknowledgments}


\vspace{5mm}
\facilities{FUSE, HST(COS)}


%\software{}
%          Cloudy \citep{2013RMxAA..49..137F}, 
%          Source Extractor \citep{1996A&AS..117..393B}
%          }

\clearpage

\appendix

\restartappendixnumbering

\section{Selected Features in the Spectrum of III-60}

\startlongtable
\begin{deluxetable*}{lcDrcc}
\tablecaption{Selected CNO Features \label{tab:lines_cno}}
\tablehead{
\colhead{Ion} & \colhead{$\lambda_{\rm lab}$} & \multicolumn2c{$\log gf$} & \colhead{$E_l$} & \colhead{Abundance} & \colhead{Grating} \\
\colhead{} & \colhead{(\AA)} & \multicolumn2c{} & \colhead{(cm$^{-1}$)}
}
\decimals
\startdata
\cthree  & 1175.665\tablenotemark{a} & 0.39 & 52419.400 & $-5.81 \pm 0.09$ & FUSE \\
              & 1175.665\tablenotemark{a} & 0.39 & 52419.400 & $-5.49 \pm 0.15$ & G130M \\
\cthree   &  1247.383  &  -0.31  &  102352.040  & $ -6.06 \pm 0.29 $ &  G130M \\
\cfour &  1107.591  &  0.03  &  320050.090  & $ -5.51 \pm 0.18$ &  FUSE \\
    	  &  1107.930  &  0.29  &  320081.692  & \nodata &  \\
 	  &  1168.847  &  0.61  &  324879.804  & $ -6.35 \pm 0.30$ & FUSE \\
	  &  1168.990  &  0.77  &  324890.316  & \nodata &  \\
 	  &  1168.847  &  0.61  &  324879.804  & $ -5.20 \pm 0.15$ & G130M \\
	  &  1168.990  &  0.77  &  324890.316  & \nodata &  \\
	  &  1548.195  &  -0.42  &  0.000  & $ -5.44 \pm 0.20 $ & G160M \\
          &  1550.772  &  -0.72  &  0.000  & \nodata  \\
\nthree & 979.876\tablenotemark{a} & 0.15 & 101027.0 & $-3.40 \pm 0.18 $  & FUSE \\ 
           & 1005.993 & -0.81 & 131004.300 & $-3.13 \pm 0.16$ & \\
           & 1006.036 & -1.12 & 131004.300 & \nodata  & \\
           & 1182.971 & $-0.924$ & 145875.700 & $-3.33 \pm 0.16$ &  \\
           & 1183.032 & $-0.608$ & 145875.700 & \nodata &  \\
           & 1184.514 & $-0.212$ & 145985.800 & \nodata &   \\
           & 1184.574 & $-0.915$ & 145985.800 & \nodata & \\
           & 1182.971 & $-0.924$ & 145875.700 & $ -3.39 \pm 0.13 $ &  G130M \\
           & 1183.032 & $-0.608$ & 145875.700 & \nodata &  \\
           & 1184.514 & $-0.212$ & 145985.800 & \nodata &   \\
           & 1184.574 & $-0.915$ & 145985.800 & \nodata & \\
\nfour  &  1131.488  &  -0.47  &  406022.802  & $ -3.33 \pm 0.16 $ & FUSE \\
  	  &  1132.021  &  -1.05  &  405971.587  & \nodata&  \\
	  &  1132.225  &  -1.17  &  405987.486  & \nodata &  \\
	  &  1132.677  &  -0.95  &  406022.802  & \nodata &  \\
	  &  1132.944  &  -1.05  &  405987.486  & \nodata &  \\
	  &  1133.121  &  -0.41  &  377284.782  & \nodata &  \\
	  &  1135.252  &  -0.63  &  377284.782  & \nodata &  \\
	  &  1136.273  &  -1.11  &  377284.782  & \nodata &  \\
           &  1195.567  &  -0.58  &  420049.619  & $ -3.37 \pm 0.24 $ & G130M  \\
  	  &  1195.732  &  -1.06  &  420049.619  & \nodata &  \\
   	  &  1195.852  &  -0.31  &  420058.003  & \nodata &  \\
            & 1718.550 & $-0.289$ & 130693.900 & $-3.27 \pm 0.13$ & G160M \\
\othree  &  1138.535  &  -0.76  &  210461.787  & $ -4.29 \pm 0.19 $ &  FUSE \\
	 &  1149.634  &  -1.08  &  197087.711  & $ -4.27 \pm 0.15 $ &  \\
	 &  1150.884  &  -0.60  &  197087.711  & \nodata &  \\
	 &  1153.775  &  -0.38  &  197087.711  & \nodata &  \\
	 &  1196.753  &  -0.35  &  294002.875  & $ -4.22 \pm 0.33 $ &  G130M \\
	 &  1197.239  &  -0.19  &  294223.069  & \nodata &  \\
	 &  1197.331  &  -0.52  &  293866.497  & \nodata &  \\
\ofour &  1046.313  &  -0.44  &  390247.987  & $ -4.28 \pm 0.54 $ & FUSE  \\
	 &  1080.967  &  0.53  &  501509.203  & $ -4.15 \pm 0.45 $ &   \\
	 &  1080.969  &  0.68  &  501564.414  & \nodata &  \\
	 &  1080.970  &  0.47  &  501509.203  & \nodata &  \\
	 &  1080.970  &  0.62  &  501564.414  & \nodata &  \\
	 &  1081.024  &  -0.94  &  390161.195  & \nodata &  \\
 	 &  1338.615  &  -0.63  &  180480.794  & $ -4.24 \pm 0.20 $ &  G130M \\
 	 &  1342.990  &  -1.33  &  180724.206  & $ $ &  \\
 	 &  1343.514  &  -0.38  &  180724.206  & $ $ &  \\
\ofive &  1371.296  &  -0.33  &  158797.700  & $ -4.21 \pm 0.25 $ &  \\
\enddata
\tablecomments{Abundance relative to hydrogen: \abund{X}.}
\tablenotetext{a}{Multiplet.}
\end{deluxetable*}

\startlongtable
\begin{deluxetable*}{lcDrcc}
\tablecaption{Additional Absorption Features \label{tab:lines_fuv}}
\tablehead{
\colhead{Ion} & \colhead{$\lambda_{\rm lab}$} & \multicolumn2c{$\log gf$} & \colhead{$E_l$} & \colhead{Abundance} & \colhead{Grating} \\
\colhead{} & \colhead{(\AA)} & \multicolumn2c{} & \colhead{(cm$^{-1}$)}
}
\decimals
\startdata
\sifour &  1066.614  &  0.72  &  160374.406  & $ -5.17 \pm 0.32 $ & FUSE \\
        &  1066.636  &  -0.59  &  160374.406  & \nodata &  \\
        &  1066.650  &  0.56  &  160375.589  & \nodata &  \\
        &  1128.325  &  -0.48  &  71748.643  & $ -5.61 \pm 0.28 $ &  \\
        &  1128.340  &  0.47  &  71748.643  & \nodata &  \\
        &  1393.755  &  0.03  &  0.000  & $ -5.11 \pm 0.18 $ &  G130M \\
        &  1402.770  &  -0.28  &  0.000  & $ -5.29 \pm 0.21  $ &  \\
\pfour    &  1030.514  &  -0.44  &  68146.475  & $ -7.29 \pm 0.18 $ & FUSE \\
          &  1030.515  &   0.25  &  68615.174  &  \nodata & \\
          &  1033.112  &  -0.32  &  68146.475  & $ -7.06 \pm 0.28 $ & \\
\pfive    &  1117.977  &  -0.01  &  0.000  & $ -6.77 \pm 0.17 $ & \\
          &  1128.008  &  -0.32  &  0.000  & $ -6.85 \pm 0.18 $ & \\
\sfour  &  1062.678  &  -1.09  &  0.000  & $ -5.09 \pm 0.42 $ & FUSE \\ 
          &  1072.996  &  -0.83  &  951.100  & $ -5.92 \pm 0.23 $ & \\
          &  1073.528  &  -1.79  &  951.100  &  \nodata  & \\
          &  1098.357  &  -1.75  &  94103.097  & $ -5.77 \pm 0.20 $ & \\
          &  1098.917  &  -0.61  &  94150.403  &  \nodata  & \\
          &  1099.472  &  -0.80  &  94103.097  &  \nodata  & \\
\sfive  &  1122.042  &  0.09  &  234956.003  & $ -5.27 \pm 0.17 $ & \\
          &  1128.667  &  -0.07  &  234947.098  &  \nodata  & \\
          &  1128.776  &  -0.97  &  234956.003  &  \nodata  & \\
          &  1268.493  &  -0.30  &  270700.417  & $ -5.57 \pm 0.35 $ & G130M \\
          &  1501.760  &  -0.50  &  127150.700  & $ -5.31 \pm 0.44 $ & G160M \\
\fefive & 1406--1412\tablenotemark{a} & \nodata &  \nodata\phn\phn  & $ -5.22 \pm 0.25 $ & G160M \\
       & 1413--1418\tablenotemark{a} & \nodata &  \nodata\phn\phn  & $ -5.38 \pm 0.27 $ \\
       & 1419--1421\tablenotemark{a} & \nodata &  \nodata\phn\phn  & $ -5.39 \pm 0.22 $ \\
       & 1428--1431\tablenotemark{a,b} & \nodata &  \nodata\phn\phn  & $ -5.31 \pm 0.25 $ \\
       & 1440--1450\tablenotemark{a} & \nodata &  \nodata\phn\phn  & $ -5.50 \pm 0.30 $ \\
       & 1450--1460\tablenotemark{a} & \nodata &  \nodata\phn\phn  & $ -5.60 \pm 0.34 $ \\
       & 1460--1470\tablenotemark{a} & \nodata &  \nodata\phn\phn  & $ -5.53 \pm 0.29 $ \\
       & 1470--1480\tablenotemark{a} & \nodata &  \nodata\phn\phn  & $ -5.38 \pm 0.36 $ \\
       & 1531--1534\tablenotemark{a,b} & \nodata &  \nodata\phn\phn  & $ -5.23 \pm 0.26 $ \\
\nifour   &  1398.193  &  0.58  &  110410.598  &  $ -6.69 \pm 0.35 $  & G130M \\
          &  1411.451  &  0.45  &  111195.796  &  $ -6.36 \pm 0.26 $  &  \\
          &  1411.451  &  0.45  &  111195.796  &  $ -6.47 \pm 0.30 $  & G160M \\
          &  1452.220  &  0.70  &  139289.405  &  $ -6.32 \pm 0.43 $  &  \\
          &  1493.672  &  0.70  &  155253.703  &  $ -6.77 \pm 0.93 $  &  \\
\nifive   &  1265--1268\tablenotemark{b}  & \nodata &  \nodata\phn\phn  & $ -6.44 \pm 0.39 $  & G130M \\
          &  1276.958  &  0.17  &  164525.901  &  $ -6.75 \pm 0.70 $  &  \\
          &  1307.603  &  0.17  &  178019.794  &  $ -6.50 \pm 0.25 $  &  \\
          &  1313.280  &  0.30  &  208163.695  &  $ -6.34 \pm 0.73 $  &  \\
\enddata
\tablecomments{Abundance relative to hydrogen: \abund{X}.}
\tablenotetext{a}{Spectral region with multiple features.  See discussion in text.}
\tablenotetext{b}{Region includes \fefour\ features.}
\end{deluxetable*}


% \bibliography{apjmnemonic,myref,stars}{}
% \bibliographystyle{aasjournal}



\end{document}